\documentclass[prd,twocolumn,amsmath,amssymb,nofootinbib,preprintnumbers,balancelastpage]{revtex4}
\usepackage{graphicx}
\usepackage{hyperref}
\hypersetup{
    pdfnewwindow=true,      
    colorlinks=true,        
    linkcolor=black,        
    citecolor=blue,         
    filecolor=blue,         
    urlcolor=blue           
}
\usepackage{graphicx}
\usepackage{cancel}
\usepackage{amssymb}
\usepackage{textcomp}
\usepackage{amsmath}
\usepackage{bm}
\usepackage{times}
\usepackage{epsfig}
\usepackage{color}
\usepackage{graphics}
\usepackage{hyperref}
\usepackage{setspace}
\usepackage{epsfig}
\usepackage{amsmath}
\usepackage{bm}
\usepackage{times}
\usepackage{graphicx}
\usepackage{color}
\usepackage{slashed}
\usepackage{graphicx}
\usepackage{amsmath, amssymb}
\usepackage{array}
\def\bea{\begin{eqnarray}}
\def\eea{\end{eqnarray}}
\begin{document}
\title{\Large {{\bf{Low Scale Unification of Gauge Interactions}}}}
\author{Pavel Fileviez P\'erez}
\author{Sebastian Ohmer}
\affiliation{\vspace{0.15cm} \\  Particle and Astro-Particle Physics Division \\
Max-Planck Institute for Nuclear Physics {\rm{(MPIK)}} \\
Saupfercheckweg 1, 69117 Heidelberg, Germany}
\date{\today}
\begin{abstract}
We investigate the possibility to realize the unification of the Standard Model gauge interactions at the low scale in four dimensions.
We find that the fields needed to define a minimal theory where baryon and lepton numbers are local symmetries, 
allow for unification at the low scale in agreement with experiments. In these scenarios the proton 
is stable and we briefly discuss the implications for cosmology. 
\end{abstract}
\maketitle
\section{Introduction}
%
Grand Unified Theories (GUTs) are considered as one of the most appealing extensions of the Standard Model 
where one understands the origin of the gauge interactions. In the context of GUTs~\cite{SU(5)} the Standard Model 
interactions are just different manifestations at low-energy of the same fundamental interaction. Unfortunately, 
these theories are realized at very high energy scales, $M_{GUT} \sim 10^{15-16}$ GeV~\cite{GQW}, and one cannot hope to test these ideas directly in experiments.
Supersymmetric unification~\cite{Dimopoulos} is very appealing and works very well in the context of the Minimal Supersymmetric Standard Model. 
However, the unification is also only possible at energy scales around $10^{16}$ GeV or even at the 
String scale, see for example~\cite{MPU}. The most generic prediction coming from grand unification 
is proton decay~\cite{review} and in most of the GUT models one needs an extra suppression 
mechanism to satisfy the experimental limits on the proton lifetime. This is the main reason 
to assume the large energy gap between the electroweak and GUT scale which is known as the great desert. 

It is well-known that in the Standard Model baryon and lepton numbers are conserved symmetries 
at the classical level. In our modern view, the Standard Model is just an effective theory which 
describes physics at the electroweak scale. In general, one should think about the impact of 
all possible higher-dimensional operators which modify the Standard Model predictions. For example, 
in the Standard Model one can have the following higher order effective operators 
\begin{eqnarray}
{\cal L} & \supset & \frac{c_L}{\Lambda_L} \ell_L \ell_L H^2  +  \frac{c_B}{\Lambda_B^2} q_L q_L q_L \ell_L \nonumber \\
& + & \frac{c_{F}}{\Lambda_F^2} (\bar{q}_L \gamma^\mu q_L) ( \bar{q}_L \gamma_\mu q_L)  
+  \ldots,
\end{eqnarray}       
where the first operator violates lepton number, the second violates baryon number and the third breaks the flavour symmetry of the Standard Model gauge sector. The experimental bounds demand 
$\Lambda_L < 10^{14}$ GeV, $\Lambda_B > 10^{15}$ GeV \cite{review}, and $\Lambda_F > 10^{3-4}$ TeV~\cite{Isidori}, when $c_L$, $c_B$ and $c_F$ are of order one. Since in any generic GUT one generates the second operator due to new gauge interactions, it is impossible to achieve unification at the low scale without predicting an unstable proton and therefore one needs the great desert.
If one could imagine a theoretical framework where the second operator is absent, the proton is stable and the unification scale is only constrained by flavor physics and has to be above $10^4$ TeV.
This is the main goal of this article.

In this article, we investigate the possibility to achieve unification at the low scale 
in the context of theories where baryon and lepton numbers are local symmetries 
spontaneously broken at the low scale. In these theories, the proton is stable and 
the new fields needed to define an anomaly free theory allow for unification at the low scale. 
One can imagine different scenarios where the unification scale could be between $10^{4}$ TeV 
and the Planck scale. If the unification scale is smaller than $10^{4}$ TeV, one needs 
an extra mechanism to suppress the flavor violating effective operators. Following the recent 
results presented in Refs.~\cite{B1,B2,B3,B4} we find that the lepto-baryons needed to cancel the 
baryonic and leptonic anomalies allow us to have unification at the low scale. 
We discuss in detail the possible scenarios where the unification 
scale is around $10^4$ TeV and point out the main features of the model.
See Fig.~1 for the unification of Standard Model gauge interactions at $10^4$ TeV.
\begin{figure}[t]
	\centering
		\includegraphics[width=0.5\textwidth]{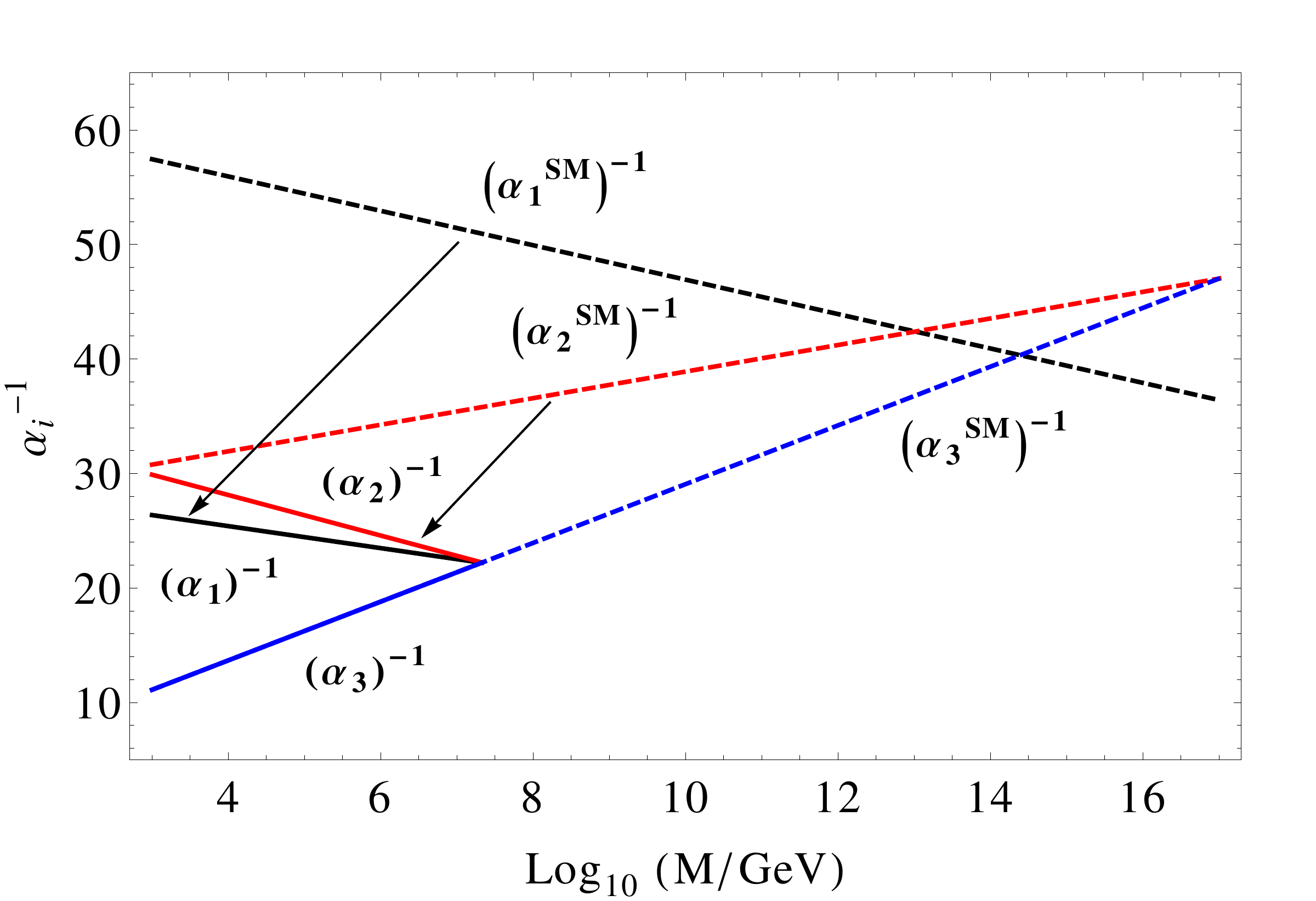}
		\caption{Evolution of the gauge couplings in the Standard Model represented by dashed lines
		and in the new model where the unification scale is realized at $10^4$ TeV. See the text for details.}
\end{figure}
%
\section{Unification of Gauge Forces}
In order to investigate the unification of gauge interactions 
at the low scale it is important to make sure that the proton 
is stable or the baryon number violating operators are highly 
suppressed. Recently, the minimal theory~\cite{B4} where the baryon and lepton numbers can be 
defined as local symmetries has been pointed out. This theory is based on the gauge 
group $$SU(3)_C \otimes SU(2)_L \otimes U(1)_Y \otimes U(1)_B \otimes U(1)_L,$$ 
and the new fermions called 
``lepto-baryons" needed for anomaly cancellation are
\begin{align}
\Psi_L \sim (1 , 2 , 1/2 , B, L ), \Psi_R \sim (1, 2, 1/2, -B, -L ), \nonumber \\
\Sigma_L \sim (1, 3, 0, -B, -L),   \  {\rm and } \  \chi_L \sim (1, 1, 0, -B, -L),     \nonumber
\end{align}
where $B=L=3/2 n_F$, and $n_F$ is the number of copies. 
Notice that these fields change only the evolution of the $SU(2)_L$ and $U(1)_Y$ gauge couplings.

The Higgs sector of this theory is composed of the fields $S_B \sim (1, 1, 0, 2B, 2L)$ 
and $S_L \sim (1, 1, 0, 0, 2)$. Therefore, when $S_L$ acquires a vacuum expectation 
value we will have $\Delta L=\pm 2$ processes. When $U(1)_B$ is spontaneously broken 
and $n_F \neq 3n$ ($n$ is an integer number) the proton is stable because one only has $\Delta B=\pm 3/n_F$ interactions.
The evolution of the Standard Model gauge couplings at one-loop level 
is described by the equation
\begin{equation}
k_i\alpha^{-1}_i(M) = \alpha^{-1}_i(M_Z) - \frac{B_i}{2 \pi} \ {\rm Log}\left(\frac{M}{M_Z}\right),
\end{equation}
where $k_i = (k_1, k_2, k_3)$ are the normalization factors. In the simplest Grand Unified Theory based on $SU(5)$ 
one has $k_i=(5/3,1,1)$. However, in general the $k_1$ value depends of the embedding in a given theory. 
The coefficients
\begin{equation}
B_i =  b^\text{SM}_i  + \theta (M-M_F) \ n_F \ b^{\rm{new}}_i  \times \frac{{\rm Log}(M/M_F)}{{\rm Log}(M/M_Z)},
\end{equation}
contain all possible contributions from the low scale, $M_Z$, to the unification scale, $M_U$.
Here, $M_F$ is the mass scale of the new fermions, $\theta (x)$ is the step function, and 
$b^{\rm{SM}}_i = (41/6, -19/6, -7)$ are the coefficients in the Standard Model.

Using the above equations one can solve for the unification scale as a function of the 
values of the gauge couplings at the $M_Z$ scale and the mass of the new fields 
\begin{equation}
{\rm{Log}} (M_U) = \frac{ C(M_Z)+n_F (b^{\rm{new}}_2-b^{\rm{new}}_3) {\rm{Log}} (M_F)}{b^\text{SM}_2-b^\text{SM}_3 + n_F (b^{\rm {new}}_2 - b^{\rm new}_3)}, 
\end{equation}
where
\begin{equation}
C (M_Z) = 2\pi (\alpha^{-1}_2(M_Z)-\alpha^{-1}_3(M_Z))+(b^\text{SM}_2 - b^\text{SM}_3) {\rm Log}(M_Z).
\end{equation}
In the same way, we can find the expression for the value of $k_1$ for given values of $M_F$
\begin{align}
k_1 = \frac{2\pi\alpha^{-1}_1(M_Z) - b^\text{SM}_1 {\rm Log} \left(\frac{M_U}{M_Z}\right) - n_F b^{\rm new}_1 {\rm Log} \left(\frac{M_U}{M_F}\right)}
{2\pi\alpha^{-1}_3(M_Z) - b^{\rm SM}_3 {\rm Log}\left(\frac{M_U}{M_Z}\right)- n_F b_3^{\rm new} {\rm Log} \left(\frac{M_U}{M_F}\right)}.
\end{align}
In the model discussed above, where baryon and lepton numbers are local symmetries, 
the new coefficients are $b_i^{\rm{new}}=(2/3,2,0)$. Now, we are ready to understand the numerical 
results assuming the unification of gauge interactions. In Table I we show different scenarios 
where $n_F$ takes the values from one to five, excluding $n_F=3$. Assuming for simplicity that the 
new fermions are at the same scale, $M_F=500$ GeV, we compute the 
unification scale and the allowed values for the $k_1$ parameter. As one can see the 
unification scale can be as low as $10^4$ TeV in agreement with the constraints from flavor violation. 
In all these scenarios the proton is stable.
\begin{table}
\begin{tabular}{lclclcl}
\hline \hline
$n_F$ & $M_U ({\rm TeV})$ & $k_1$ \\
\hline \hline
$1$ & $1.24 \cdot 10^{9}$ & $2.05$ \\ 
$2$ & $4.96 \cdot 10^6$ & $2.67$ \\ 
$4$ & $2.14 \cdot 10^4$ & $3.62$ \\ 
$5$ & $4.58 \cdot 10^3$ & $3.99$ \\ 
\hline \hline
\end{tabular}
\caption{Solutions for the unification at the low scale in models with gauged baryon and lepton numbers.}
\end{table}

In Fig.~1 we show the evolution of the gauge couplings in the Standard Model and in the model with lepto-baryons. The lepto-baryons at the low scale change the evolution of $\alpha_2$ and $\alpha_1$ dramatically without affecting $\alpha_3$. We assume four copies of lepto-baryons with baryon and lepton numbers equal to $3/8$. Since they have different baryon and lepton numbers than the Standard Model fields we never induce new sources of flavor violation and the unification scale $M_U$ is larger than $10^4$ TeV in order to suppress the flavor violating effective operators discussed in the introduction. 

The lightest neutral field in the new sector is always stable after symmetry breaking as discussed in Ref.~\cite{B4} and one has a candidate to describe the cold dark matter in the Universe. The collider bounds on the new fermions are weak because they can only be produced through electroweak interactions at the Large Hadron Collider (LHC). Therefore, this type of model is in agreement with all constraints coming from collider experiments and cosmology.   
\subsection{Baryonic and Leptonic Couplings}
In this minimal theory for local baryon and lepton numbers 
the evolution of the new couplings is defined by the equation
\begin{align}
\alpha^{-1}_X (M_Z) &= k_X \alpha^{-1}_X (M_U) + \frac{B_X}{2 \pi} \ {\rm Log}\left(\frac{M_U}{M_Z}\right),
\end{align}
where $X=B,L$. The $B_X$ coefficients are given by
\begin{align}
B_B = b^\text{SM}_B + \theta (M_U-M_F) \ b^{\rm{new}}_B \times \frac{{\rm Log}(M_U/M_F)}{{\rm Log}(M_U/M_Z)},
\end{align}
and
\begin{align}
B_L = b^\text{SM}_L + \theta (M_U-M_F) \ b^{\rm{new}}_L \times  \frac{{\rm Log}(M_U/M_F)}{ {\rm Log}(M_U/M_Z)} \,.
\end{align}
Notice that the values of these gauge couplings at different scales and the $k_X$ are barely constrained. However, one can envision that it is possible to define a theory where all gauge couplings are unified, the Standard Model couplings, $\alpha_1$, $\alpha_2$, $\alpha_3$, and the new couplings $\alpha_B$ and $\alpha_L$.Therefore, assuming unification at the $M_U$ scale, we can find the values of $\alpha_B$ and $\alpha_L$ at the $M_Z$ scale for given values of $k_B$ and $k_L$.
The contribution of the Standard Model fields to the running of the baryonic and 
leptonic couplings is given by the coefficients $b^\text{SM}_B = 8/3$ and $b^\text{SM}_L = 6$.
The new fields contribute as follows
\begin{align}
b^{\rm{new}}_B &= \left(\frac{16}{3}n_F + \frac{4}{3}\right)B^2={3} \frac{(4 n_F + 1)}{n_F^2}, \\
b^{\rm{new}}_L &=  b^{\rm{new}}_B + \frac{10}{3} \,.
\end{align}
Notice that the right-handed neutrinos and $S_L$ contribute to the running of the leptonic coupling, 
defining the difference between the running of $\alpha_B$ and $\alpha_L$. 
\begin{figure}[ht]
	\centering
		\includegraphics[width=0.5\textwidth]{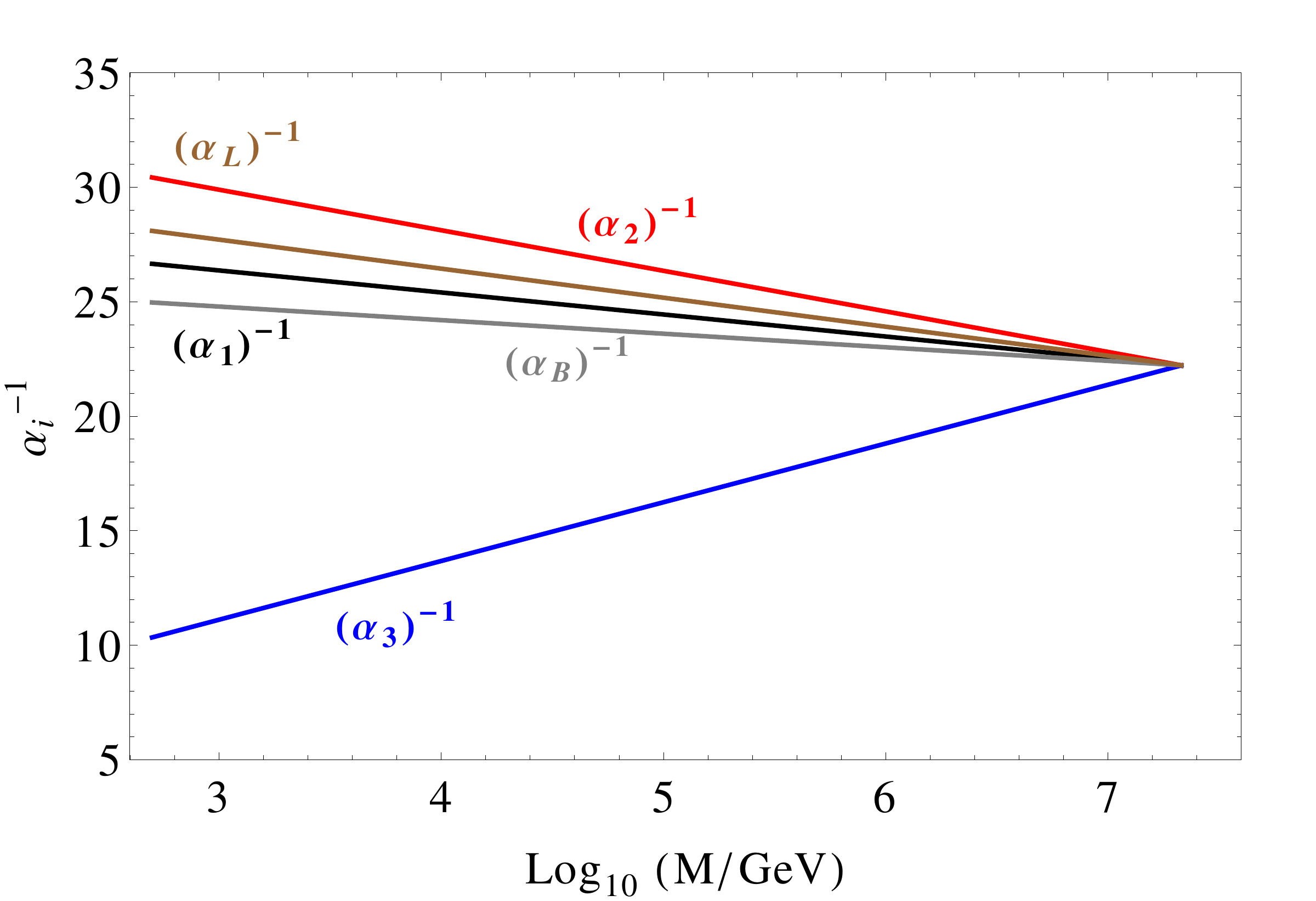}
	\caption{Evolution of the gauge couplings with lepto-baryons where the unification scale is  $M_U \sim 2 \times 10^4$ TeV. Here we assume for simplicity $k_B=k_L=k_1$ for the running of the $\alpha_B$ and $\alpha_L$. At the $M_Z$ scale we find $\alpha^{-1}_B (M_Z) = 25.17$ and $\alpha^{-1}_L (M_Z) = 28.69$.}
\end{figure}
Hence, we can calculate the values of these couplings assuming unification at the low scale.
For simplicity, we have assumed the same $k_i$ factor for all Abelian symmetries in the theory.

In Fig.~2 we show the numerical results for the running of all gauge couplings. 
As one can appreciate the couplings $\alpha_B$ and $\alpha_L$ are small 
at the electroweak scale. This result is 
welcome because in this way one can satisfy the bounds on the 
leptophobic $Z_B$ and quarkphobic $Z_L$ gauge couplings coming 
from collider experiments. See Refs.~\cite{Dobrescu,An} for the bounds 
on the leptophobic gauge bosons. 

Now, let us explicitly write down the relevant interactions for the model 
in order to understand how the different fields can obtain mass after symmetry 
breaking. The relevant interactions are given by
\begin{eqnarray}
-\mathcal{L}  & \supset  & h_1  \bar{\Psi}_R  H \chi_L + h_2  H^\dagger \Psi_L \chi_L 
+  h_3  H^\dagger \Sigma_L  \Psi_L  \nonumber \\ 
&+& h_4  \bar{\Psi}_R  \Sigma_L H +  \lambda_\Psi \ \bar{\Psi}_R \Psi_L S_B^* + \lambda_\chi \ \chi_L \chi_L S_B \nonumber \\
&+ & \lambda_{\Sigma} \ \text{Tr} \ \Sigma_L^2 S_B + Y_\nu \ \ell_L H \nu^c + \lambda_R  \  \nu^c \nu^c S_L \ + \   \text{h.c.}, \nonumber \\
\end{eqnarray}
where $\nu^c=(\nu_R)^c$ are the right-handed neutrinos and $H \sim (1, 2, 1/2, 0, 0)$ 
is the Standard Model Higgs boson. In general for $n_F$ copies of the fields the couplings 
$h_1$, $h_2$, $h_3$, $h_4$, $\lambda_\Psi$, $\lambda_\chi$, and $\lambda_{\Sigma}$ 
are $n_F \times n_F$ matrices. Notice that when $S_B$ breaks $U(1)_B$ all the new 
fields can have large masses even before the electroweak symmetry is broken. 
The leptonic $U(1)_L$ symmetry is broken by the vacuum expectation value of $S_L$, and 
in the same way we generate large right-handed neutrino masses.  

In Ref.~\cite{B4} we have investigated the possible implications 
for baryogenesis if the local baryon number is broken at the low scale 
for $n_F=1$. In this scenario the sphalerons must satisfy 
the extra condition that total baryon number is conserved. 
As one can see in Ref.~\cite{B4} it is possible to have a consistent relation 
between the final baryon asymmetry and the $B-L$ asymmetry generated 
by some mechanisms in the early Universe such as leptogenesis. These results can be easily 
generalized for several copies of the new fermions.
In Ref.~\cite{B4} we also discussed the existence of a dark matter candidate which is the lightest 
Majorana fermion in the new sector. If we have several copies of these fields, 
the same candidate can be used to describe the cold dark matter in the Universe.   

Before we summarize our main results, it is important to discuss the impact 
on the running of the gauge couplings in the context of other models where 
the cancellation of the baryonic and leptonic anomalies is possible. 
In Ref.~\cite{B1} the anomaly cancellation was realized using an extra 
chiral family. Unfortunately, this solution is ruled out by the LHC due to 
the existence of a Standard Model-like Higgs. In Ref.~\cite{B2} the possibility to use vector-like families for anomaly cancellation was pointed out. These vector-like families change the running of the gauge couplings, but the unification is always realized at high scales because the new quarks change the evolution of the strong coupling as well. 

The possibility to cancel the B and L anomalies using lepto-baryons was also
discussed in Ref.~\cite{B3}. In this scenario, the new fields do 
not change the evolution of the strong coupling and 
they are singlets or live in the fundamental representation 
of $SU(2)$. Since the fundamental representation of 
$SU(2)$ contributes less than the adjoint representation 
one will need many copies of these fields to achieve 
the low scale unification. Therefore, we focus on the 
scenarios studied in this article since they are the minimalistic setting where we can make sure that 
the proton is stable after symmetry breaking and the 
unification can be realized at the low scale as shown in Fig.~1 and Fig.~2.
We have presented our results at one-loop level in order to illustrate 
the main idea. It is also important to understand the predictions 
at two-loop level and in a given unified theory the threshold effects could be very important.   

\section{Concluding Remarks}
\label{summary}
In this article, we have shown that the unification of the Standard Model 
gauge interactions can be achieved at the low scale. In order to investigate this 
issue we focused on theories where baryon and lepton numbers are 
defined as local gauge symmetries spontaneously broken at the low scale.
In this class of theories, the proton is stable and the unification of the gauge 
couplings can be realized at the low scale in a consistent way. 
We have illustrated the numerical results in Fig.~1 and Fig.~2 
for the case $n_F=4$, where the unification scale 
is $M_U \sim 2 \times 10^4$ TeV, in agreement with the 
constraints from flavor violating decays. In our opinion, one can 
have an even lower unification scale adding more copies 
of the new fermions, but it is only reasonable
if there is a mechanism to suppress flavor violation.
We listed one example in Table I where the unification 
is below $10^4$ TeV in order to complete our studies.  

The low unification scale which can be achieved with lepto-baryons motivates 
the possibility to test the idea of grand unification at future colliders. One should say that these results are just one step towards 
the possibility to have a complete unified theory at the low scale.
We have identified which type of theories allow us to realize low scale unification. 
In this context, using different copies of lepto-baryons needed for anomaly 
cancellation we find the allowed values for the $k_i$ coefficients for exact unification.
These results should motivate the model builders in the GUT and String communities 
since now one can search for alternative ways to have unification at the low scale.
In our opinion, these results tell us that the dream of testing the 
unification of gauge interactions is maybe not too far. 
The main priority in this field is to find the simplest unified theory which 
could allow us to have the embedding of this low energy theory 
with gauged baryon and lepton numbers. 

{\textit{Acknowledgments:}}
{\small {We thank Pran Nath, Hiren H. Patel and Mark B. Wise for discussions. }}



\end{document}